\documentclass[twocolumn,aps,prb,superscriptaddress,floatfix]{revtex4-2}

\usepackage[english]{babel}
\usepackage{times}
\usepackage{graphicx}
\usepackage{graphics}
\usepackage{amsmath}
\usepackage{amsfonts}
\usepackage{amssymb}
\usepackage{epstopdf}
\usepackage{makeidx}
\usepackage{subfigure}
\usepackage{color}
\usepackage{pgf}
\usepackage{bm}
\usepackage{tikz} 
\usepackage[normalem]{ulem}
\usepackage{soul}
\usepackage{hyperref}
\usepackage{float}
\usepackage{ulem}

\newcommand{\angstrom}{\text{\normalfont\AA}}

\newcommand{\be}{\begin{equation}}
\newcommand{\ee}{\end{equation}}
\newcommand{\bea}{\begin{eqnarray}}
\newcommand{\eea}{\end{eqnarray}}

\definecolor{byzantine}{rgb}{0.56, 0.0, 1.0}

\makeindex
\begin{document}

\title{Probing topological phase transitions via quantum reflection in the graphene family materials} 
\author{P. P. Abrantes}
\email{ppabrantes91@gmail.com}
\affiliation{Instituto de F\'{\i}sica, Universidade Federal do Rio de Janeiro, 21941-972, RJ, Brazil}
\author{Tarik P. Cysne}
\email{tarik.cysne@gmail.com}
\affiliation{Instituto de F\'{\i}sica, Universidade Federal Fluminense, 24210-346, RJ, Brazil}
\author{D. Szilard}
\affiliation{Instituto de F\'{\i}sica, Universidade Federal do Rio de Janeiro, 21941-972, RJ, Brazil}
\author{F. S. S. Rosa}
\affiliation{Instituto de F\'{\i}sica, Universidade Federal do Rio de Janeiro, 21941-972, RJ, Brazil}
\author{F. A. Pinheiro}
\affiliation{Instituto de F\'{\i}sica, Universidade Federal do Rio de Janeiro, 21941-972, RJ, Brazil}
\author{C. Farina}
\affiliation{Instituto de F\'{\i}sica, Universidade Federal do Rio de Janeiro, 21941-972, RJ, Brazil}
%
%

\begin{abstract}

We theoretically investigate the quantum reflection of different atoms by two-dimensional (2D) materials of the graphene family (silicene, germanene, and stanene), subjected to an external electric field and circularly polarized light. By using Lifshitz theory to compute the Casimir-Polder potential, which ensures that our predictions apply to all regimes of atom-2D surface distances, we demonstrate that the quantum reflection probability exhibits distinctive, unambiguous signatures of topological phase transitions that occur in 2D materials. We also show that the quantum reflection probability can be highly tunable by these external agents, depending on the atom-surface combination, reaching a variation of 40\% for Rubidium in the presence of a stanene sheet. Our findings attest that not only dispersive forces play a crucial role in quantum reflection, but also that the topological phase transitions of the graphene family materials can be comprehensively and efficiently probed via atom-surface interactions at the nanoscale.

\end{abstract}

\maketitle


\section{Introduction}

Two-dimensional (2D) materials constitute a remarkable material platform to investigate fluctuation-induced phenomena, at the interdisciplinary frontiers between condensed matter, atomic and molecular physics, and materials science \cite{Woods-RPM, Klimchitskaya-2009, Laliotis-Wilkowski-2021, BSLu-2021}. It was theoretically predicted that dispersive forces [{\it e.g.} Casimir and Casimir-Polder (CP) interactions] in graphene-based systems can be efficiently controlled by external means, such as the application of magnetic fields \cite{Macdonald_PRL, Cysne-Kort-Kamp_2014, QR-MagField-2019}, strain engineering \cite{Nichols-2016}, carriers doping \cite{Cysne-2016, Bimonte-2017, Bordag-2016}, and suitable stacking in multilayer 2D systems \cite{Khusnutdinov-2016, Abbas-2017}. Other quantum vacuum-related effects (as, for instance, CP torque in anisotropic molecules \cite{Antezza-2020} and quantum friction \cite{Belen-1, Belen-3}) have also been explored in 2D systems. Significant experimental progress in the investigation of such phenomena in 2D materials has also been made \cite{Sulimany-2018, Banishev-2013, Klimchitskaya-experiment, Liu-Mohideen-experiment_2}, specially in the case of graphene. As a result, it is clear that 2D materials play a pivotal role in the current and future understanding of Casimir interactions and related physical phenomena \cite{grapheneFamily_Nature}.

Among different 2D materials, the so-called graphene family materials (silicene, germanene, and stanene - the 2D allotropes of Si, Ge, and Sn, respectively) single out for exhibiting fascinating physical phenomena and a broad range of applications \cite{Castellanos_2016}. They were synthesized only recently \cite{Stanene-syntesis, Germanene-syntesis, Silicene-syntesis} and, in contrast to graphene, which has a spin-orbit coupling (SOC) around a few $\mu$eV, the intrinsic SOC in these materials is higher \cite{Ezawa_2015, Liu_Jian_Yao_2011, Matthes_2013}, a direct effect of their larger atomic number when compared to carbon. Their intrinsic SOC, ranging from $2$ to $50$ meV, leads to a robust quantum spin Hall insulator (QSHI) phase characterized by a $\mathbb{Z}_2$-topological invariant \cite{KM-Z2}. These materials possess a buckled honeycomb lattice and, as a result, topological phase transitions occur due to the characteristic response of their electronic structures under the application of external electric bias \cite{Ezawa_2013}.

Quantum reflection (QR) consists of the reflection of incident quantum particles under the influence of a potential that decreases monotonically in the direction of the particle motion, despite the absence of any turning points. It finds practical applications in the field of atom-optics \cite{Cronin-RMP-2009, Deutschmann-1993, Landragin-1997, Shimizu-Fujita-2002, Kohno-2003, Savalli-2002}, such as in the design of atomic mirrors \cite{Judd-2010, Judd-2011, Segev-1997, Aspect-1996}, atomic traps \cite{Trap-Crepin-2017, Jurisch-2008}, and diffraction gratings \cite{Keith-1988, Zhao-2008} and it is also related to the study of interactions involving Bose-Einstein condensates \cite{Pasquini-2004,Pasquini-2006,Wang-Xiong-BEC}. Furthermore, QR has been employed as an efficient method to probe dispersive interactions between atoms and surfaces \cite{Friedrich-2002, Shimizu-2001, Zhao-2008, Bender-2010, Druzhinina-2003, Barnea-2017} (see Ref. \cite{Bai-2019} for application in atomic collisions). The understanding of such interactions involving surfaces of complex materials \cite{Woods-RPM, Dalvit-2008, Intravaia-2020} is crucial for the description of many relevant phenomena in chemistry \cite{Ernst-2021} and applied physics \cite{Bender-2014, Barnea-2017} of current interest.

In this paper, we investigate the QR of different atomic species by representative 2D materials of the graphene family. Topological phase transitions are induced by applying an external electric field and by irradiating circularly polarized light in the 2D sheets. We show that QR probability can be significantly modified by changing the intensities of these external agents, specially in the case of materials with high intrinsic SOC and more massive atoms. Our results demonstrate that topological phase transitions clearly show up in the behavior of QR as a function of the external electric field, suggesting that QR could be used as an alternative and effective optical method to probe topological phase transitions in 2D materials and other topologically non-trivial systems.

This paper is organized as follows. In the next section, we present the methodology employed to evaluate the QR probabilities. Section \ref{sec3} comprises our results for the QR probability of a rubidium atom by different materials of the graphene family and Sec. \ref{sec4} is dedicated to our final comments and conclusions. Appendixes \ref{appendA} and \ref{appendB} contain important information on the mathematical description of materials and atoms studied in this work, whereas Appendix \ref{appendC} presents results for QR of a sodium atom.


\section{Methodology} \label{sec2}

\begin{figure}[h!]

	\centering
	\includegraphics[width=0.98\linewidth,clip]{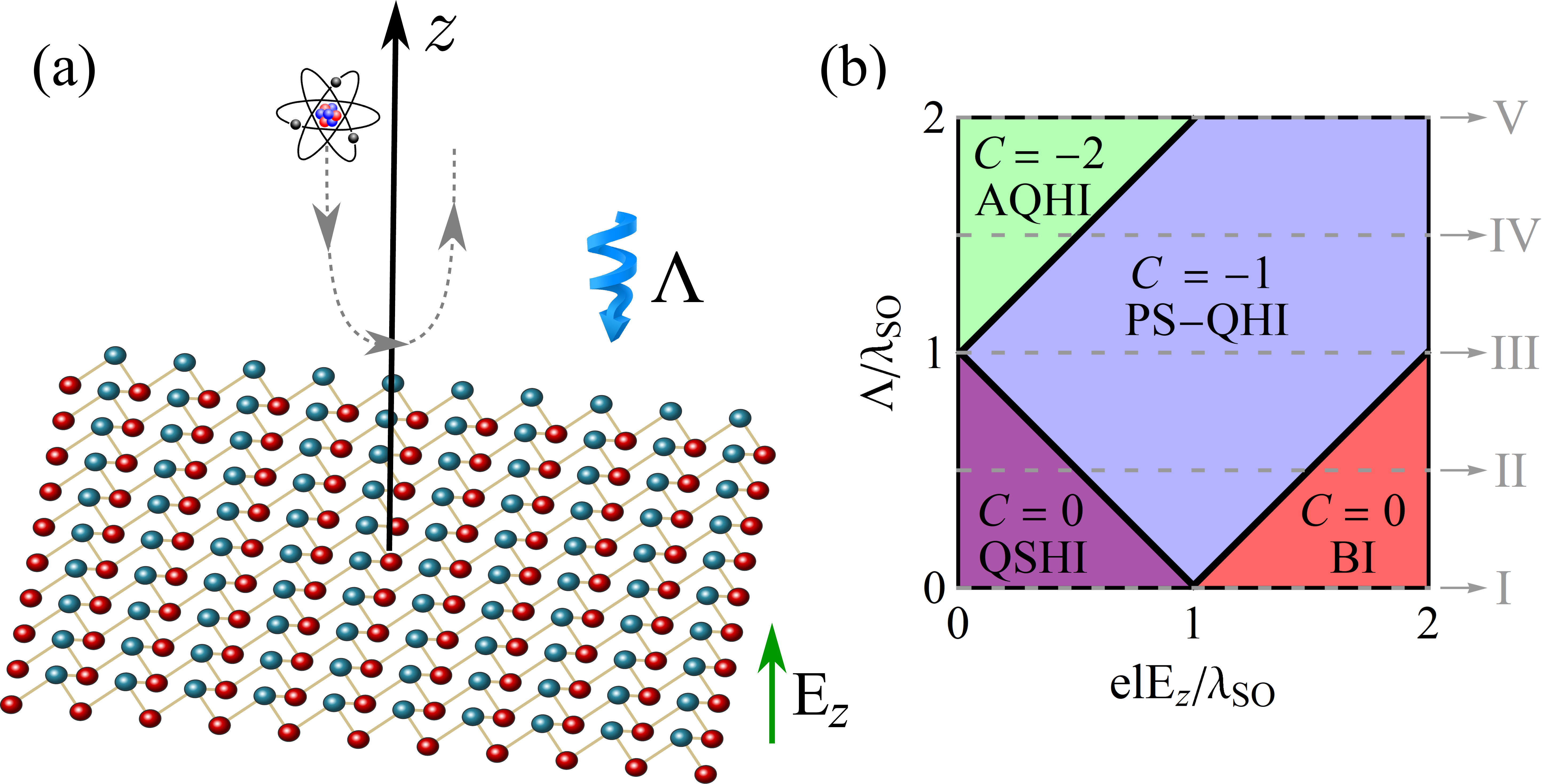}
	\caption{\textbf{(a)} Atomic specimen being reflected by a suspended 2D material of the graphene family. The system is under the influence of a static electric field $E_z$ and it is shined by a circularly polarized light, characterized by the parameter $\Lambda$. \textbf{(b)} Topological phase diagram of the 2D graphene family materials. Horizontal dashed gray lines show the paths used in this work to explore this diagram. The acronyms of each topological phase means: band insulator (BI; $C = 0$), quantum spin Hall insulator (QSHI; $C = 0$, indexed by the non-trivial $\mathbb{Z}_2$-index \cite{KM-Z2}), polarized-spin quantum Hall insulator (PS-QHI; $C~=~-~1$) and anomalous quantum Hall insulator (AQHI; $C=-2$), with $C$ standing for the corresponding Chern numbers.}
	\label{Fig.QR}

\end{figure}

Figure \ref{Fig.QR}(a) depicts the physical situation of interest. An atomic beam moves towards a suspended sheet of a 2D graphene family material (silicene, germanene, or stanene) with normal incidence. Each atom with energy $E$ interacts with the attractive potential generated by the 2D surface and, due to the quantum particles wave nature, it has a non-zero probability to be reflected by this attractive potential. An electric field $E_z$ is perpendicularly applied to the 2D material which is also irradiated by a circularly polarized light, characterized by the parameter $\Lambda$ (see Appendix \ref{appendB} for the precise definition of this parameter). Figure \ref{Fig.QR}(b) illustrates the topological phase diagram in the space of dimensionless parameters $e \ell E_z/\lambda_{\textrm{SO}}$ and $\Lambda/\lambda_{\textrm{SO}}$, where $\lambda_{\textrm{SO}}$ is the spin-orbit coupling, $e$ is the modulus of electron charge and $\ell$ is the material buckling. A description of this diagram can be found in Appendix \ref{appendB}. For further details, see Ref. \cite{Ezawa_2013}.

The attractive potential between incident atomic specimens and a given 2D suspended material of the graphene family is determined by the Casimir-Polder (CP) potential. At low temperatures, the CP potential between a neutral but polarizable atom and a flat surface can be calculated using the Lifshitz formula \cite{Antezza-2020, Lifshitz-1956, Messina-2009} (see also Refs. \cite{Fialkovsky-2018, Buhmann-Marachevsky-Scheel} for a thorough derivation of the Lifshitz formula in materials with finite Hall conductivity), according to
\begin{align}
	U (z) &= \frac{\hbar}{\epsilon_0 c^2} \int_0^{\infty} \! \frac{d\xi}{2\pi} \xi^2 \,\alpha(i\xi) \int \!\! \frac{d^2{\bm k}}{(2\pi)^2} \frac{e^{-2 \kappa z}}{2 \kappa}  \cr\cr
	& \times \left[ r^{ss} ({\bm k}, i\xi) - \left( 1 + \frac{2 c^2 k^2}{\xi^2} \right) r^{pp} ({\bm k}, i\xi) \right] \,,
\label{u}
\end{align}
where $\kappa = \sqrt{\xi^2/c^2 + k^2}$, $\alpha(i \xi)$ denotes the atomic electric polarizability as a function of imaginary frequencies $i \xi$, and $r^{ss} ({\bm k}, i\xi)$ and $r^{pp} ({\bm k}, i\xi)$ stand for the diagonal reflection coefficients associated to the 2D material. As usual, $s$ and $p$ mean the transverse electric and transverse magnetic polarizations, respectively. Note that, in contrast to configurations involving only macroscopic objects, the reflection coefficients mixing the polarizations $s$ and $p$ do not contribute to Eq. (\ref{u}) when one considers the interaction between an isotropic atom and an macroscopic object with Hall conductivity \cite{Fialkovsky-2018, Rosa-2008}. The reflection coefficients, the model for the atomic polarizability, and other parameters related to each atom considered in this work can be found in Appendix \ref{appendA}. For the sake of simplicity, we shall neglect thermal corrections and the roughness of the 2D material in this present work \cite{Bezerra-2008, Buhmann-Scheel-2008, Alves-2021}.

Considering an atom of mass $m$ and energy $E$ under the influence of any potential $U (z)$, the time-independent Schr\"odinger equation reads
\begin{eqnarray}
	\frac{\partial^2 \psi (z)}{\partial z^2} + \frac{p^2(z)}{\hbar^2} \psi (z) = 0 \,,
\label{SE}
\end{eqnarray}
with
\begin{equation}
	p (z) = \sqrt{2 m [E - U (z)]} \,.
\label{p}
\end{equation}
The WKB solutions are good approximations when the atom is located far from the monolayer, when compared to the length scale associated to the CP interaction, to wit, $c/\xi_l$ (see Table \ref{table1} in Appendix \ref{appendA}). In such a case, a solution of the form \cite{Berry-1972}
\begin{eqnarray}
	\psi (z) = \frac{c_+ (z)}{\sqrt{|p(z)|}} e^{i \phi(z)} + \frac{c_- (z)}{\sqrt{|p(z)|}} e^{-i \phi(z)} \,, 
\label{WKB}
\end{eqnarray}
is a very convenient one, where $\phi (z)$ is written as
\begin{eqnarray}
	\phi (z) = \int^{z}_{z_0} \! dz'\, \frac{p(z')}{\hbar} \,.
\label{WKBphi}
\end{eqnarray} 
The previous ansatz is very suitable, since it transforms the second order Schr\"odinger equation into a set of two coupled first-order differential equations for the coefficients $c_+ (z)$ and $c_- (z)$. In fact, substituting Eqs. (\ref{WKB}) and (\ref{WKBphi}) into Eq. (\ref{SE}), it can be shown that \cite{Dufour-2015-tesis}
\begin{eqnarray}
	\frac{\partial c_+ (z)}{\partial z} &=& e^{- 2 i \phi (z)} \frac{c_- (z)}{2 p(z)} \frac{\partial p(z)}{\partial z} \,,
\label{DE1} \\ \cr
	\frac{\partial c_- (z)}{\partial z} &=& e^{2 i \phi (z)} \frac{c_+ (z)}{2 p(z)} \frac{\partial p(z)}{\partial z} \,.
\label{DE2}
\end{eqnarray}
It is reasonable to assume that any atom that reaches the surface of the material will not be reflected, but adsorbed to it,  leading to the boundary conditions $c_+ (0) = 0$ and $c_- (0) = 1$.  
The quantum reflection probability is then defined as \cite{Dufour-2013, Dufour-2-2013, Dufour-2015-tesis, Judd-2011,Crepin-2017}
\begin{eqnarray}
	R = \lim_{z \rightarrow \infty} \Bigg| \frac{c_+ (z)}{c_- (z)} \Bigg|^2 \,. 
\label{QRP}
\end{eqnarray}

Information about the efficiency of a given potential $U (z)$ to give rise to QR can be extracted from the so-called badlands function \cite{Dufour-2013, Dufour-2-2013, Dufour-2015-tesis, Judd-2011,Crepin-2017}
\begin{eqnarray}
	Q (z) = \frac{\hbar^2}{2 p^2 (z)} \left[ \frac{\phi''' (z)}{\phi' (z)} - \frac{3}{2} \left( \frac{\phi'' (z)}{\phi' (z)}\right)^2 \right] \,,
\label{QF}
\end{eqnarray}
with $p(z)$ and $\phi(z)$ given by (\ref{p}) and (\ref{WKBphi}) and the primes indicate derivatives with respect to $z$. It is common in the literature to associate the highest probabilities of QR occurrence to regions of highest values of $Q(z)$ \cite{Dufour-2013, Dufour-2-2013, Dufour-2015-tesis, Judd-2011,Crepin-2017}. However, for very low energies (very high de Broglie wavelengths), this is not the case and QR must be thought of as a quite non-local effect so that the correlation between the QR probability and the badlands function seems to be not valid anymore. In other words, the unsuitability of this function only occurs at the threshold of the QR ($E \rightarrow 0$) \cite{Petersen2018}. Nevertheless, as will be discussed in the next section, this work is concerned with the intermediate energy regime, where tuning the QR probability by applying external fields is more easily achieved, therefore, the usual interpretation based on badlands functions still holds. For a given energy, the badlands function exhibits a peak whose maximum occurs, by assumption, at  position $z_m$. To solve Eqs. (\ref{DE1}) and (\ref{DE2}) numerically, we need to choose points $z_i$ and $z_f$ of the space, such that $z_i \ll z_m \ll z_f$. The boundary conditions are applied at $z_i$, {\it i.e.}, $c_+ (z_i) = 0$ and $c_- (z_i) = 1$, and the limit of Eq. (\ref{QRP}) is taken at $z_f$, {\it i.e.}, $R = |c_+ (z_f)/c_- (z_f)|^2$. In this sense, $z_i$ and $z_f$ can be understood as convergence parameters and their typical values are around a few Angstroms and a fraction of centimeters, respectively. For further details on the methods of solving coupled differential equations in the QR problem, we refer the reader to Refs. \cite{Dufour-2013, Dufour-2-2013, Dufour-2015-tesis, Judd-2011,Crepin-2017}.

We computed the QR probability for two atomic species of experimental relevance (Rb \cite{Marchant-2016} and Na \cite{Pasquini-2004, Pasquini-2006}), and for three graphene family materials already synthesized (stanene \cite{Stanene-syntesis}, germanene \cite{Germanene-syntesis}, and silicene \cite{Silicene-syntesis}) in vacuum. For a given atomic specie, with parameters specified in Table \ref{table1}, and a given reflective material, whose parameters are presented in Table \ref{table2}, the QR probability is a function of the energy of the incident particles beam, the applied electric field, and the intensity of the circularly polarized laser. By modifying these external agents, the 2D materials undergo different topological phase transitions that drastically affect their optical conductivities (see Appendix \ref{appendB}) and, consequently, the atomic reflection probability. We found that the effect of external perturbations in QR is more pronounced for heaviest atoms. In what follows, we present our results for QR of a Rb atom by the graphene family materials. Similar results for a Na atom can be found in Appendix \ref{appendC}.


\section{Results and discussions} \label{sec3}

\begin{figure}[b!]

	\centering
	\includegraphics[width=0.91\linewidth,clip]{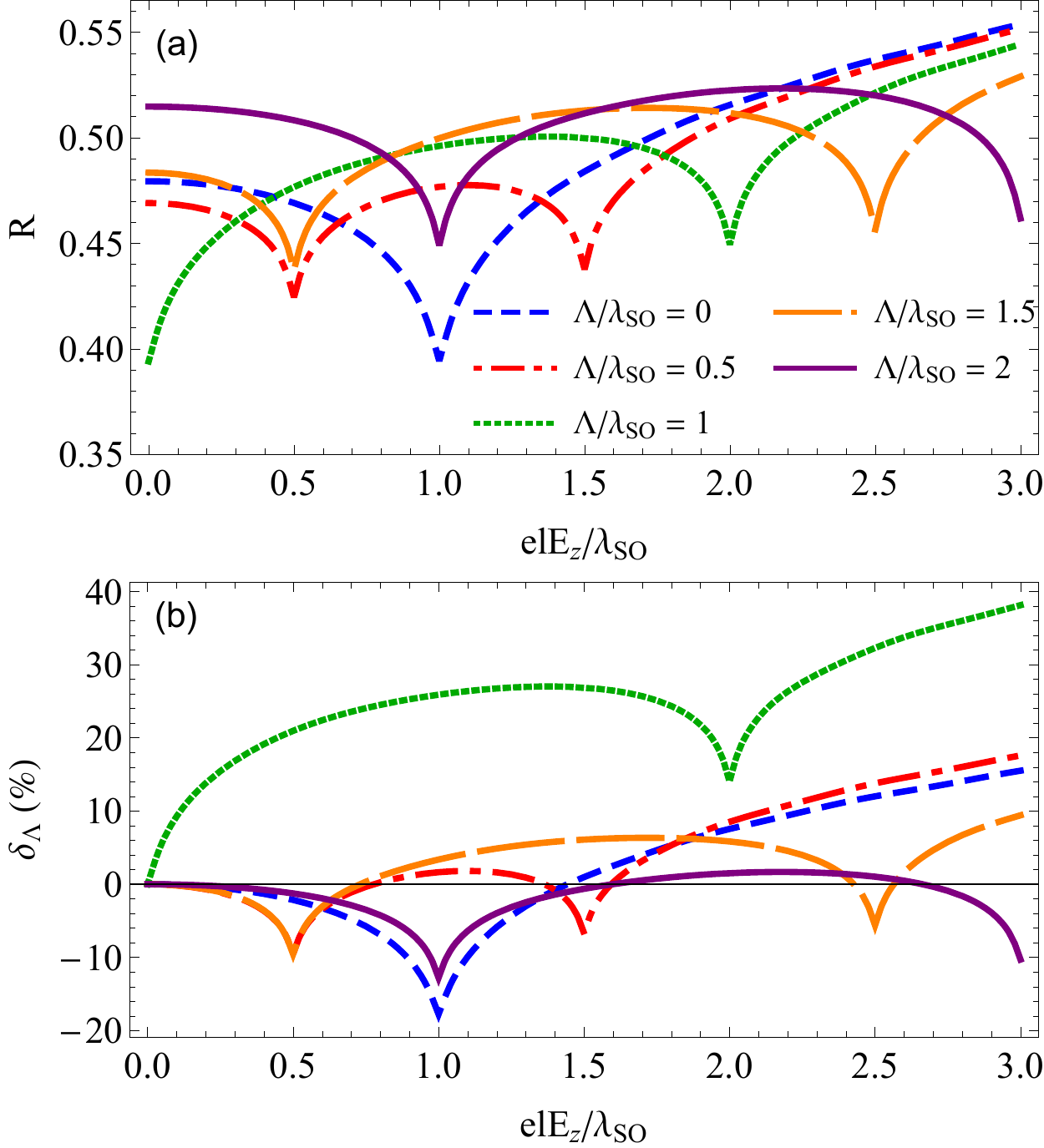}
	\caption{Results for QR of a Rb atom with energy $E = 10^{-4}$~neV by a suspended sheet of pristine stanene ($\lambda_{\textrm{SO}} = 50$ meV, chemical potential $\mu = 0$ and inverse of the scattering time $\Gamma = 10^{-4} \lambda_{\textrm{SO}}/\hbar$, as described in Appendix \ref{appendB}). \textbf{(a)} QR probability as a function of the electric field for different values of the laser parameter. \textbf{(b)} Percentual modification in the QR probability caused by the electric field.}
	\label{FigureMain1}

\end{figure}

Figure \ref{FigureMain1} exposes the results for QR of a Rb atom by a stanene sheet under the influence of electric field and circularly polarized light. We consider that the energy of the incident Rb atom is $E = 10^{-4}$ neV (see next discussion). In Fig. \ref{FigureMain1}(a), we show the QR probability as a function of the applied electric field for five values of the laser parameter $\Lambda$. Each curve corresponds to a path symbolized by a horizontal dashed gray line in Fig. \ref{Fig.QR}(b). Let us start with the analysis of the blue curve, with $\Lambda/\lambda_{\textrm{SO}} = 0$, meaning that the circularly polarized light is absent [path I in Fig. \ref{Fig.QR}(b)] and the electric field varying from $e \ell E_z/\lambda_{\textrm{SO}} = 0$ to $e \ell E_z/\lambda_{\textrm{SO}} = 3$. The key feature that we highlight is the non-monotonic behavior of the QR probability. It starts decreasing from $R \approx 0.48$ until it reaches the minimum value of $R \approx 0.39$ at the critical point of the topological phase transition $e \ell E_z/\lambda_{\textrm{SO}} = 1$. At this point, the electronic spectrum of stanene becomes gapless and this point separates a quantum spin-Hall insulator (QSHI) phase from a trivial band insulator (BI) one. If we continue to increase the intensity of the electric field going into the deep trivial insulator phase, the QR probability increases and reaches $R \approx 0.55$ for $e \ell E_z/\lambda_{\textrm{SO}} = 3$. It is worth mentioning that both topological phases discussed above are indexed by zero Chern number, but this is not synonymous of topological triviality. In fact, it is possible to attribute other topological invariant to the quantum spin Hall insulator phase called $\mathbb{Z}_2$ invariant, which is not trivial \cite{KM-Z2}. But this $\mathbb{Z}_2$ invariant does not have a direct connection with charge conductivity as it is the case of Chern number (see Appendix \ref{appendB}).

Now, we consider the presence of circularly polarized light, turning on the laser parameter to a constant value of $\Lambda/\lambda_{\textrm{SO}}~=~0.5$, represented by the red curve in Fig. \ref{FigureMain1}(a). In such a configuration, the cusp of the later case splits into others, stamping the existence of two topological phase transitions. This is in agreement with the topological phases crossed by path II in Fig. \ref{Fig.QR}(b), in which there is an emergence of a third topological phase called polarized-spin quantum Hall insulator (PS-QHI), between $e \ell E_z/\lambda_{\textrm{SO}} = 0.5$ and $e \ell E_z/\lambda_{\textrm{SO}} = 1.5$, that separates the aforementioned QSHI and BI phases. Analogously, the phase diagram of Fig. \ref{Fig.QR}(b) can be fully explored by the paths III, IV, and V associated with $\Lambda/\lambda_{\textrm{SO}} = 1, 1.5, 2$, respectively, and their results for the QR probability are also presented in Fig. \ref{FigureMain1}(a). A remarkable characteristic of all curves is the presence of a cusp exactly at the topological phase transition points. This suggests at least two routes to explore these results. Firstly, one can take advantage of these phase transitions to control and tune the QR probability, adjusting these external agents (the circularly polarized laser and the electric field) to enhance it or diminish it according to what may be more convenient. Secondly, these results may provide a great opportunity of using the QR as a simple optical tool to probe these topological phase transitions experimentally, since the QR probability changes its behavior whenever the combination of electric field and laser intensities hit a point of the phase transition diagram.

\begin{figure}[b!]

	\centering
	\includegraphics[width=0.90\linewidth,clip]{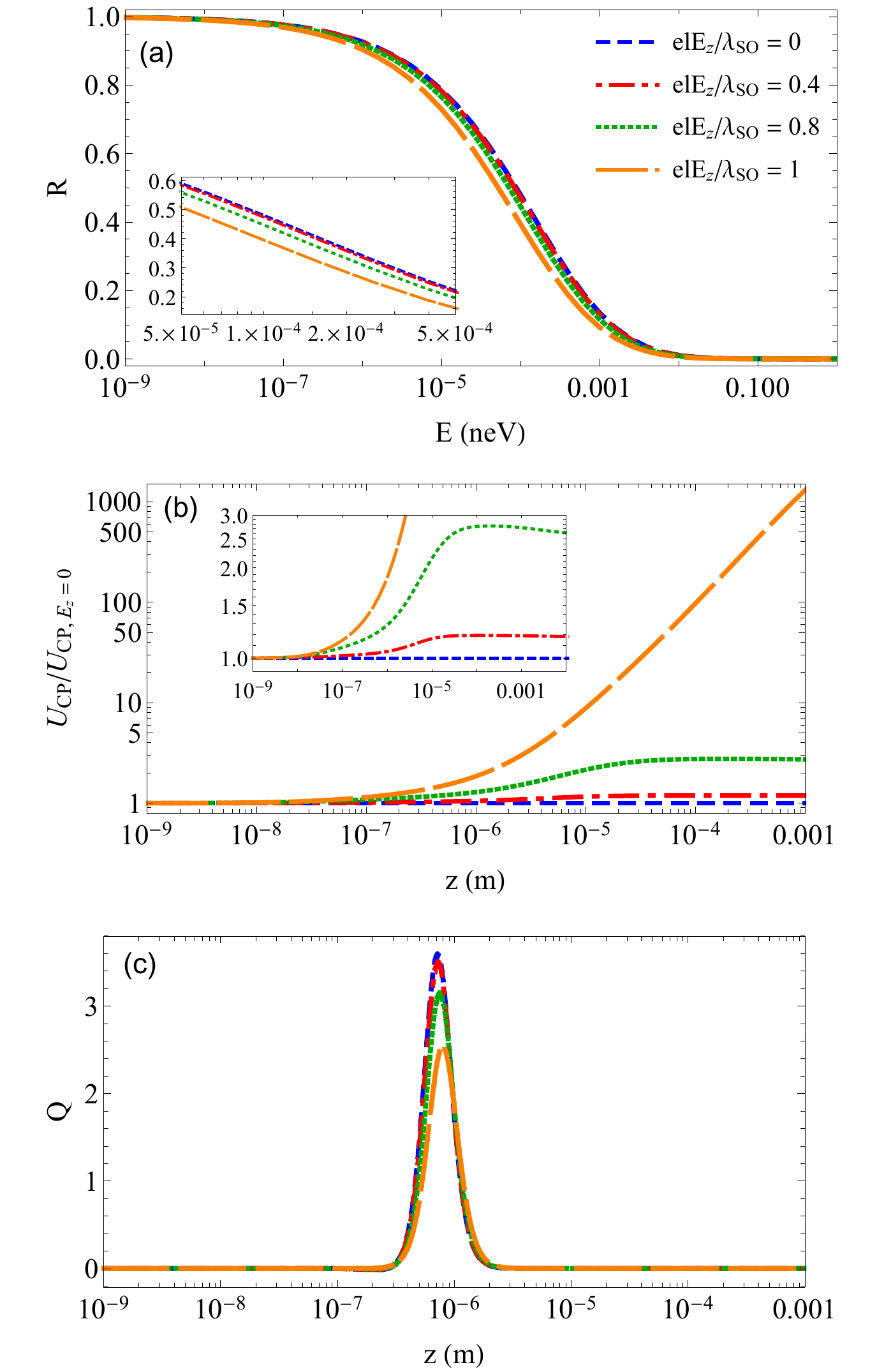}
	\caption{\textbf{(a)} The probability $R$ of a Rb atom suffering QR from a stanene sheet as a function of its incident energy $E$. \textbf{(b)} Relative modification in the CP energy between the Rb atom and the 2D surface caused by the applied electric field as a function of the distance $z$. \textbf{(c)} The plot of $Q (z)$ as a function of the distance $z$ for the incident energy $E = 10^{-4}$ neV chosen in the intermediate energy regime between the classical and the quantum limit of QR [see panel (a)]. In all plots, $\Gamma = 10^{-4} \lambda_{\textrm{SO}}/\hbar$, $\mu = 0$, and $\Lambda = 0$.}
	\label{FigApp2}

\end{figure}

\begin{figure}[b!]

	\centering
	\includegraphics[width=0.90\linewidth,clip]{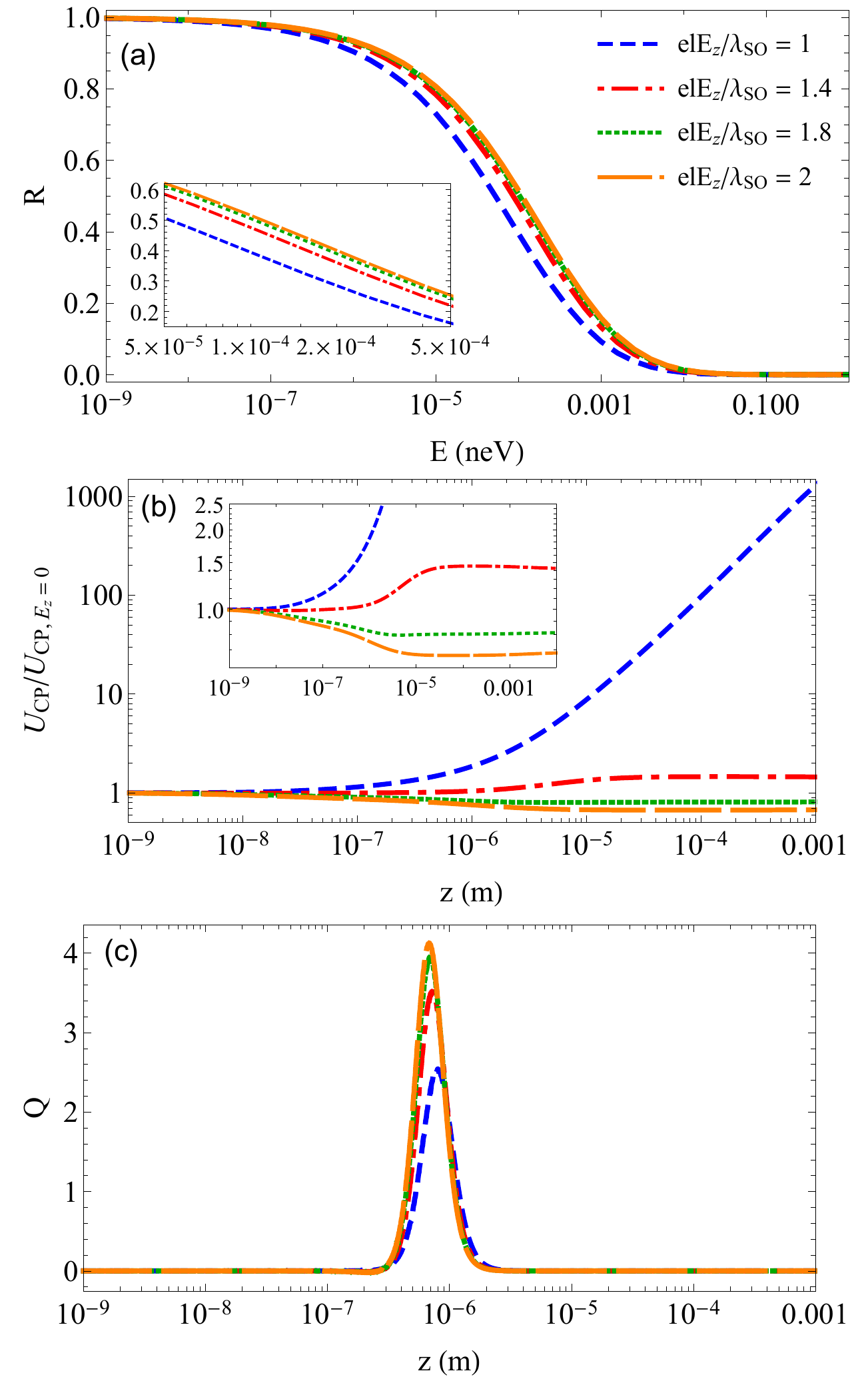}
	\caption{{\bf (a)} The probability $R$ of a Rb atom suffering QR from a stanene sheet as a function of its incident energy $E$. \textbf{(b)} Relative modification in the CP energy between the Rb atom and the 2D surface caused by the applied electric field as a function of the distance $z$. \textbf{(c)} The plot of $Q (z)$ as a function of the distance $z$ for the incident energy $E = 10^{-4}$ neV chosen in the intermediate energy regime between the classical and the quantum limit of QR [see panel (a)]. In all plots, $\Gamma = 10^{-4} \lambda_{\textrm{SO}}/\hbar$, $\mu = 0$, and $\Lambda = 0$.}
	\label{FigApp3}

\end{figure}

To achieve a deeper and more complete understanding of the phenomenology involved in the QR physics by these topological materials, we present a detailed discussion of the case where $\Lambda/\lambda_{\textrm{SO}} = 0$ [path I depicted in Fig. \ref{Fig.QR}(b)]. In Fig. \ref{FigApp2} we show the QR as a function of the incident energy [panel (a)], the normalized CP energy as a function of distance [panel (b)], and $Q (z)$ as a function of $z$ for an incident energy $E = 10^{-4}$ neV [panel (c)]. Each curve denotes a distinct value of $e \ell E_z/\lambda_{\textrm{SO}} \leq 1$, judiciously chosen to explore the region in which stanene crosses the QSHI phase up to the critical point at $e \ell E_z/\lambda_{\textrm{SO}} = 1$. In Fig. \ref{FigApp3} we present similar results but for electric field values for which stanene goes from this critical point to the BI phase ($e \ell E_z/\lambda_{\textrm{SO}} \geq 1$).

Both Figs. \ref{FigApp2} and \ref{FigApp3} clarify the cusps in QR probability associated to the topological phase transitions. From Fig. \ref{FigApp2}, it should be noticed that the QR probability presents three regimes. {\it (i)} The low energy limit, in which the wavelike nature of the quantum particle stands out and $R \rightarrow 1$ as $E \rightarrow 0$. {\it (ii)} The high energy limit, in which $R \rightarrow 0$ as $E \rightarrow \infty$ and the particle behaves classically. {\it (iii)} The intermediate regime lying in between the classical and the quantum ones where $0 < R < 1$, in which a choice of incident energy is advantageous, since the QR probability can be substantially altered by combined effects of external agents \cite{Cysne-Kort-Kamp_2014, QR-MagField-2019}. In this intermediate regime, Fig. \ref{FigApp2}(a) points out that the QR probability decreases with the electric field in the QSHI-phase for most range of energies of the incident particle. This feature can be explained with the aid of Figs. \ref{FigApp2}(b) and (c), that show the effect of electric field on the CP energy and on function $Q (z)$, respectively, for a particle with energy $E = 10^{-4}$ neV. The peak of the $Q (z)$ is located around $z \approx 1$ $\mu$m for all values of $e \ell E_z$ in the QSHI phase, meaning that this is the distance regime where the QR is most salient. In this same topological phase, the result of the electric field on the CP energy is to enhance its intensity as the system approaches the critical point [\ref{FigApp2}(b)]. This effect is noticeable for distances $z \geq 0.1$ $\mu$m only, but that contemplates the peak of $Q (z)$. As a consequence, the height of the peak of $Q (z)$ decreases and the outcome is a decrease of the QR probability when the electric field is varied from $e \ell E_z/\lambda_{\textrm{SO}} = 0$ to $1$ [\ref{FigApp2}(c)]. An opposite behavior can be identified when we consider the BI phase presented in Fig. \ref{FigApp3}. As can be verified in panel (a), the QR probability increases with the intensity of the electric field in intermediate energy regimes. In addition, panel (b) reveals that the CP energy decreases with the electric field at long distances, which causes the height of the peak of $Q (z)$ to increase in panel (c). The final result translates into an enhancement of the QR probability with the increase of the electric field intensity in the BI phase. Ultimately, the results presented in Figs. \ref{FigApp2} and \ref{FigApp3} can be traced back to the cusp at the topological critical point in the curve of $\Lambda = 0$ in Fig. \ref{FigureMain1}(a). This same reasoning can be borrowed to every results of the main text in order to understand all structures of cusps.

We can also discuss the degree of tunability of the QR probability. To this end, we define the quantity
\begin{eqnarray}
	\delta_{\Lambda} (E_z) = \left[ \frac{R_{\Lambda} \left( E_z \right) - R_{\Lambda}\left( E_z = 0 \right)}{R_{\Lambda} \left( E_z = 0 \right)} \right] \times 100\% \,,
\label{percentual}
\end{eqnarray}
that furnishes an estimate of the QR relative variation induced by the applied electric field for a given laser intensity. Figure \ref{FigureMain1}(b) shows the degree of tunability of the QR probability for the same parameters employed in Fig. \ref{FigureMain1}(a), revealing an amplitude of variation that comfortably lies in the measurement precision of QR experiments \cite{Marchant-2016, Pasquini-2004, Pasquini-2006, Nayak-1983, Berkhout-1989, Doyle-1991, Savalli-2002}.

\begin{figure}[t!]
	\centering
	\includegraphics[width=0.913\linewidth,clip]{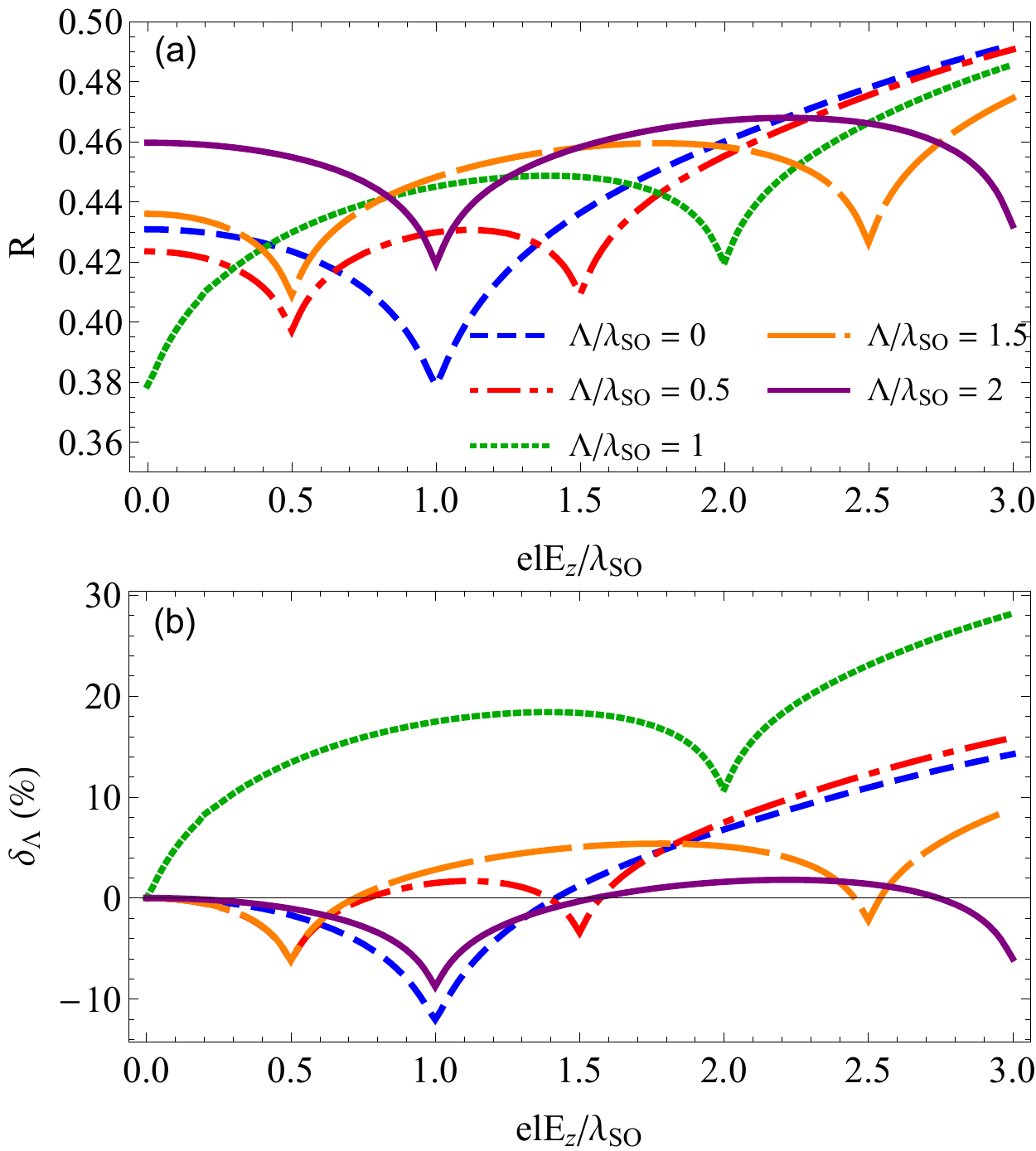}
	\caption{Results for QR of a Rb atom with energy $E = 10^{-4}$~neV by a suspended sheet of pristine germanene ($\lambda_{\textrm{SO}} = 20$ meV, $\mu = 0$ and $\Gamma = 10^{-4} \lambda_{\textrm{SO}}/\hbar$). \textbf{(a)} QR probability as a function of the electric field for different values of the laser parameter. \textbf{(b)} Percentual modification in the QR probability caused by the electric field.}
	\label{FigureMain2}
\end{figure}

\begin{figure}[t!]

	\centering
	\includegraphics[width=0.913\linewidth,clip]{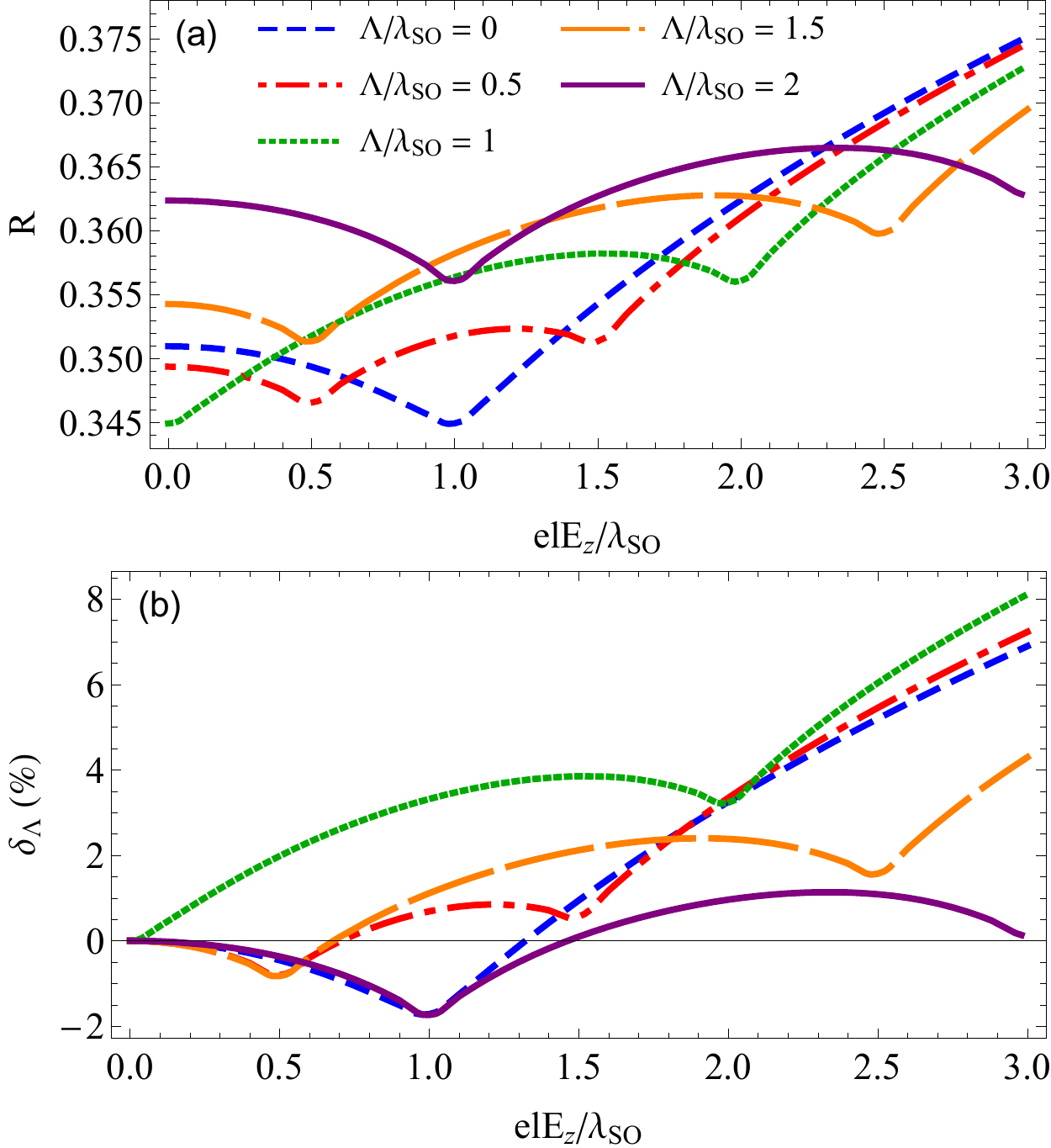}
	\caption{Results for QR of a Rb atom with energy $E = 10^{-4}$~neV by a suspended sheet of pristine silicene ($\lambda_{\textrm{SO}} = 2$ meV, $\mu = 0$ and $\Gamma = 10^{-4} \lambda_{\textrm{SO}}/\hbar$). \textbf{(a)} QR probability as a function of the electric field for different values of the laser parameter. \textbf{(b)} Percentual modification in the QR probability caused by the electric field.}
	\label{FigureMain3} 

\end{figure}

Figures \ref{FigureMain2} and \ref{FigureMain3} show similar results, but for QR of a Rb atom by the other two materials (germanene and silicene, respectively), indicating that QR by germanene is also quite sensitive. Unfortunately, due to the lower spin-orbit coupling of silicene compared with the ones of germanene and stanene, the control of QR by the electric field is not so intense in this material, as can be seen contrasting Fig. \ref{FigureMain3}(b) with Figs. \ref{FigureMain1}(b) and \ref{FigureMain2}(b). Despite that, the noticeable presence of cusp when the material crosses a given topological phase transition persists in the case of silicene and their optical detection could also remain possible. From the perspective of material synthesis, it is noteworthy to mention that it should be easier to fabricate germanene and silicene with currently available methods \cite{Stanene-syntesis, Germanene-syntesis, Silicene-syntesis}.

As a final comment, we briefly discuss the last cousin of the graphene family, the plumbene, synthesized in 2019 \cite{Plumbene-syntesis}. Although its spin-orbit coupling is almost twice as strong as that one of stanene, density-functional calculations have shown that plumbene is a band insulator (BI) at neutrality ($\mu = 0$) \cite{Plumbene-BI}. The induction of a quantum spin-Hall insulator phase can be performed by electron doping, originating a physics like the one described here \cite{Plumbene-BI}. In such a case, we expect that the effects of electric field and circularly polarized laser should be much more noticeable, as well as the control of QR probability. However, the simplified Dirac Hamiltonian described in Eq. (\ref{Hd}) and the related topological phase diagram of Fig. \ref{Fig.QR}(b) may not be applicable for plumbene, and more detailed modeling would be necessary.


\section{Final Remarks and Conclusions} \label{sec4}

We have investigated the quantum reflection (QR) of atoms by graphene family materials (stanene, germanene, and silicene). In our setup, we have considered two atomic species of experimental importance (Na and Rb) and the system was subject to a perpendicular electric field and a circularly polarized light. These external agents may induce several topological phase transitions and the fundamental motivation was to understand how they may affect atom-surface interactions. We have demonstrated that, whenever the 2D material undergoes a given topological phase transition, it leaves a clear signature in the QR probability. Therefore, our findings reveal that one can look at them from two different perspectives. The first one involves the use of these materials as a promising platform to manipulate and tune the QR probability through an external agent. The second one is to take advantage of the QR as a simple optical tool to probe these topological phase transitions experimentally. Bearing all this in mind, we also determined the sensitivity of the QR probability due to variations of the electric field and laser intensities, verifying that they lie within the scope of typical experimental precision in many situations. A higher degree of control is achieved in the case of Rb-stanene, in which the modification on the QR probability may reach 40\%. This can be mainly associated with two aspects: {\it (i)} The heaviest mass of the Rb atom, which makes the QR to be dominated by the longer distances regime of the CP energy (a region most affected by external agents), and {\it (ii)} the higher spin-orbit coupling of stanene, making the effect of perturbations more pronounced. Altogether we expect that these results allow for an alternative way to control QR in increasingly smaller scales, as well as contribute to a better understanding of the rich physics in Casimir effect and other related phenomena present in these materials.


\section*{Acknowledgements}

C.F. and F.S.S.R. acknowledge Conselho Nacional de Desenvolvimento Cient\'{i}fico e Tecnol\'{o}gico (CNPq) for financial support (Grant numbers 310365/2018-0 9 and 309622/2018-2). F.S.S.R. (Grant number E26/203.300/2017) and P.P.A. acknowledge Fundação de Amparo à Pesquisa do
Estado do Rio de Janeiro (FAPERJ). T.P.C acknowledges Coordenação de Aperfeiçoamento de Pessoal de Nível Superior (CAPES). D.S. and F.A.P also thank the funding agencies.

\appendix

\section{Reflection coefficients and atomic polarizabilities \label{appendA}}

The reflection coefficients of a 2D material sheet with finite longitudinal and transverse optical conductivities, needed in Eq. (\ref{u}) to evaluate the Casimir-Polder energy, can be obtained by solving Maxwell's equations with the appropriate boundary conditions \cite{Macdonald_PRL, Cysne-Kort-Kamp_2014, QR-MagField-2019, Catarina-2020}. These formulas can be written as
\begin{align}
	r^{ss} ({\bm k}, i\xi) &= \frac{2 \sigma_{xx} (i\xi) Z^{\textrm{H}} + \eta_0^2 [\sigma_{xx}^2 (i\xi) + \sigma_{xy}^2 (i\xi)]}{-\Phi({\bm k}, i\xi)}\,,
\label{rss} \\
	r^{pp} ({\bm k}, i\xi) &= \frac{2 \sigma_{xx} (i\xi) Z^{\textrm{E}} + \eta_0^2 [\sigma_{xx}^2 (i\xi) + \sigma_{xy}^2 (i\xi)]}{\Phi({\bm k}, i\xi)} \,,
\label{rpp} \\
	\Phi ({\bm k}, i\xi) &= [2 + Z^{\textrm{H}} \sigma_{xx}(i\xi)] [2 + Z^{\textrm{E}} \sigma_{xx} (i\xi)] \nonumber \\
	& + [\eta_0 \sigma_{xy} (i\xi)]^2  \,,
\label{RefCoefs}
\end{align}
where $Z^{\textrm{H}} = \xi \mu_0 / \kappa$, $Z^{\textrm{E}} = \kappa/(\xi \epsilon_0)$, and $\eta_0^2=\mu_0/\epsilon_0$. Additionally, $\sigma_{xx} (i\xi)$ and $\sigma_{xy} (i\xi)$ denote the longitudinal and transverse conductivities of the 2D material, respectively, as a function of imaginary frequencies $i \xi$.

In order to describe the atomic polarizabilities in the Casimir-Polder potential [Eq. (\ref{u})], we employed a Lorentz oscillator model with a single resonance, given by
\begin{eqnarray}
	\alpha_l (i\xi) = \frac{\alpha_l (0)}{1 + \frac{\xi^2}{\xi_l^2}} \,.
\label{alpha}
\end{eqnarray}

\begin{table}[b!]

	\centering
	\begin{tabular}{||c c c c c c c||} 
		\hline
		$l$ & & $\alpha_l (0)$ (a.u.) & &  $\hbar \xi_l$ (eV) & & $m (\times 10^{-27}\text{ Kg})$ \\ [0.5ex] 
 		\hline
		\hline
		Na & & 162.6 & & 2.13 & & 38.17\\
		Rb & & 318.6 & & 1.68 & & 141.92 \\ [0.5ex] 
		\hline
	\end{tabular}
	\caption{Data for Na and Rb atoms. This table contains parameters of the single Lorentz-oscillator model needed in Eq. (\ref{alpha}) \cite{Khusnutdinov-2016} and their masses. ($1$ a.u. $= 1.648 \times 10^{-41}$ C$^2$m$^2$J$^{-1}$).}
	\label{table1}

\end{table}

\noindent The different fitted parameters for each atomic specimen analyzed here are well-known from the literature, see Table \ref{table1}. For materials with higher spin-orbit coupling, such as stanene, stronger electric fields are necessary to explore the entire topological phase diagram of Fig. \ref{Fig.QR} (b). In turn, these electric fields may cause a significant modification in the atomic resonance frequency $\xi_l$ via the Stark effect. For instance, in the case of QR of Rb atom by stanene, this correction can be of the order of 25$\%$ of the values presented in Table \ref{table1}. Nevertheless, such modification only implies in a minor numerical correction in the short-distance regime of the CP energy and, consequently, does not affect our results on the QR probability. Therefore, in the numerical results presented in this work, we neglected the effects of the applied electric field and circularly polarized light on the atom, due to the dominant impact of these external agents in the optical conductivity of the reflective 2D surface (see Appendix \ref{appendB}).


\section{Optical conductivities \label{appendB}}

The low energy description of buckled materials of the graphene family under the influence of external perpendicular electric field ($E_z$) and circularly polarized laser beam ($\Lambda$) is given by the Hamiltonian \cite{Ezawa_2013},
\begin{eqnarray}
	H_s^{\eta} = \hbar v_F (\eta k_x \tau_x + k_y \tau_y) + \frac{\Delta^{\eta}_s}{2} \tau_z - \mu \,,
\label{Hd}
\end{eqnarray}
where
\begin{eqnarray}
	\Delta_s^{\eta} = \eta s \lambda_{\textrm{SO}} - e \ell E_z - \eta \Lambda \,.
\label{md}
\end{eqnarray}
In these equations, $\eta = \pm 1$ is the quantum number related to $K$ and $K'$ valleys of the Brillouin zone of the honeycomb material, $\tau_{x,y,z}$ are Pauli matrices related to the sublattice degree of freedom, $s = \pm 1$ for electron spins $\uparrow$ and $\downarrow$, and $k_{x (y)}$ are the electron momentum relative to $K$ ($K'$) valleys. Also, $v_F = \sqrt{3} a t/2$ is the Fermi velocity, $a$ is the lattice parameter, $t$ is the hopping parameter, and $\mu$ is the chemical potential of the material, which we set $\mu = 0$ in the numerical calculations of this work (see Refs. \cite{Cysne-2016, Bimonte-2017, Bordag-2016} for details about the impact of the chemical potential in the Casimir-Polder interaction in Dirac materials). In Eq. (\ref{md}), $\lambda_{\textrm{SO}}$ is the intrinsic spin-orbit coupling, $e$ is the modulus of electron charge, $\ell E_z$ is the sublattice potential generated by the external electric field $E_z$, and the parameter $\Lambda = \pm 8 \pi \alpha v_F^2 I_0/\omega_0^3$ is associated to the circularly polarized light of intensity $I_0$ and frequency $\omega_0$ irradiated in the material. Furthermore, $\alpha$ is the fine structure constant and the $+$ ($-$) sign denotes the left (right) circular polarization \cite{grapheneFamily_Wilton_PRL}. Material parameters can be found in Table \ref{table2}.

\begin{table}[h!]

	\centering
	\begin{tabular}{||c c c c c c c c c||} 
		\hline
		Material & & $\lambda_{\textrm{SO}}$ (meV) & &  $t$ (eV) & &  $\ell$ ($\angstrom$) & &  $a$ ($\angstrom$) \\ [0.5ex] 
		\hline
		\hline
		Silicene & & 2 & & 1.6 & & 0.11 & & 3.86\\
		Germanene & & 20 & & 1.3 & & 0.16 & & 4.02 \\
		Stanene & & 50 & & 1.3 & & 0.20 & & 4.70 \\ [0.5ex] 
		\hline
	\end{tabular}
	\caption{Spin-orbit coupling ($\lambda_{\textrm{SO}}$), hopping parameter ($t$), buckling ($\ell$) and lattice parameter ($a$) of each graphene family material used in this work. Parameters were taken from references \cite{Ezawa_2015, Liu_Jian_Yao_2011, Matthes_2013, Castellanos_2016}. {\it Note:} There exists some discordance in the literature for $\lambda_{\textrm{SO}}$ in the case of stanene.}
	\label{table2}

\end{table}

\begin{figure}[h!]

	\centering
	\includegraphics[width=0.99\linewidth,clip]{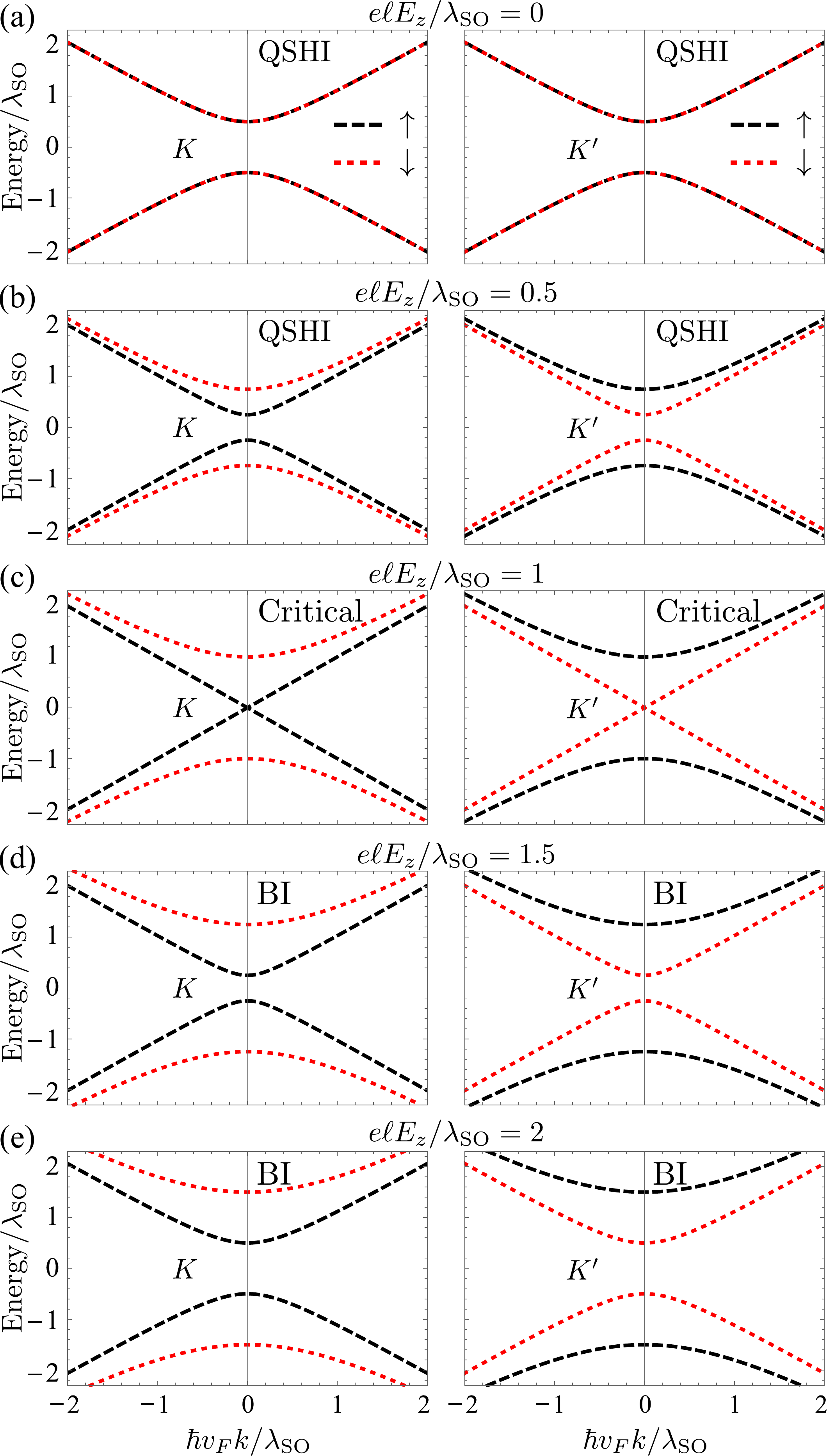}
	\caption{Evolution of the electronic spectrum of the Hamiltonian of Eqs. (\ref{Hd}) and (\ref{md}), along the path I of Fig. \ref{Fig.QR} (b), {\it i. e.}, $\Lambda/\lambda_{\rm SO} = 0$, and {\bf (a)} $e\ell E_z/\lambda_{\rm SO} = 0$, {\bf (b)} $e\ell E_z/\lambda_{\rm SO} = 0.5$, {\bf (c)} $e\ell E_z/\lambda_{\rm SO} = 1$, {\bf (d)} $e\ell E_z/\lambda_{\rm SO} = 1.5$, and {\bf (e)} $e\ell E_z/\lambda_{\rm SO} = 2$. The left panels show the spectrum for the valley $K$ ($\eta=+1$) and the right panels show the spectrum for the valley $K'$ ($\eta=-1$). The black-dashed line depicts the spectrum for the spin-$\uparrow$ sector ($s=+1$), and the red-dotted line depicts the spectrum for the spin-$\downarrow$ sector ($s=-1$). The topological critical point is represented in panel (c), where the electronic spectrum becomes gapless, and it separates the QSHI phase [panels (a) and (b)] and the BI phase [panels (d) and (e)].}
	\label{FigureSpec}

\end{figure}

In Fig. \ref{FigureSpec}, we analyze the evolution of the electronic spectrum of the Hamiltonian shown in Eqs. (\ref{Hd}) and (\ref{md}), exploring the path I ($\Lambda/\lambda_{\rm SO} = 0$) of Fig. \ref{Fig.QR} (b). For the cases presented in panels \ref{FigureSpec} (a) and \ref{FigureSpec} (b), where $e \ell E_z/\lambda_{\rm SO} = 0$ and $e \ell E_z/\lambda_{\rm SO} = 0.5$, respectively, the 2D material exhibits a QSHI phase, indexed by a non-trivial $\mathbb{Z}_2$-topological invariant, if the chemical potential $\mu$ lies inside the gap of the electronic spectrum. This topological index remains invariant over small changes in the Hamiltonian, as long as the electronic gap remains open and still contains the chemical potential \cite{KM-Z2}. However, as the electric field is enhanced, the topological gap becomes increasingly small. When the field reaches the value corresponding to $e\ell E_z/\lambda_{\rm SO} = 1$, shown in panel \ref{FigureSpec} (c), the spin-$\uparrow$ ($\downarrow$) sector of the electronic spectrum becomes gapless at the valley $K$ ($K'$), and the system touches the topological critical point. If one continues to increase the electric field intensity, the electronic gap reopens but now exhibiting a trivial-insulator (or BI) behavior, as shown in panels \ref{FigureSpec} (d) and \ref{FigureSpec} (e). The two insulating phases presented in Fig. \ref{FigureSpec} are characterized by a Chern number $C = 0$.

This previous analysis of the electronic spectrum can be repeated to understand all topological phase transitions crossed by paths II, III, IV, and V of Fig. \ref{Fig.QR} (b). For instance, by changing the parameters along path II ($\Lambda/\lambda_{\rm SO}= 0.5$), the 2D material undergoes three distinct topological phases, that are separated by two topological critical points. The first one occurs for $e\ell E_z/\lambda_{\rm SO} = 0.5$, where the spin-$\uparrow$ sector of the spectrum becomes gapless at valley $K$, while the spectrum remains gapped at valley $K'$, and separates the QSHI from the PS-QHI phase, indexed by a non-trivial Chern number $C = - 1$. The second critical point occurs for $e\ell E_z/\lambda_{\rm SO} = 1.5$, where the spin-$\downarrow$ sector of the electronic spectrum becomes gapless at valley $K'$, while it remains gapped at valley $K$, and separates the PS-QHI from the BI phase. Lastly, to analyze the phase transition between the AQHI phase, indexed by a Chern number $C = - 2$, and the PS-QHI phase, we consider path IV ($\Lambda/\lambda_{\rm SO} = 1.5$). At the critical point $e\ell E_z/\lambda_{\rm SO} = 0.5$, that separates these two topological phases, the spin-$\uparrow$ sector of the electronic spectrum becomes gapless at valley $K'$ and remains gapped at valley $K$. Furthermore, extrapolating the path IV until $e\ell E_z/\lambda_{\rm SO} = 2.5$, there is another critical point to be crossed, separating the PS-QHI and the BI phases, where the spin-$\downarrow$ sector of the spectrum becomes gapless at valley $K'$, while it remains gapped at valley $K$. This analysis exhausts the different types of phase transitions presented in Fig. \ref{Fig.QR} (b), that occur whenever the material hits the points of the parameter space for which the electronic spectrum becomes gapless. Among the topological invariants described above, the Chern number has particular importance in the context of this work, due to its connection with the DC limit of the Hall conductivity (see next paragraph) that appears in the reflection coefficients of the 2D material. The non-zero Chern numbers are related to the existence of localized edge-states in finite samples \cite{Tse-Niu-2011}, responsible for the transport of the Hall conductivity that is robust against disorder and weak interactions, a characteristic aspect of non-trivial topological systems \cite{KM-Z2, Tse-Niu-2011}.

Longitudinal and transverse optical conductivities for graphene family materials described by Hamiltonian (\ref{Hd}) can be obtained by the usual Kubo formalism in the linear response regime \cite{Stille-2012, Shah2020} and closed analytical expressions were obtained in Refs. \cite{grapheneFamily_Nature, grapheneFamily_Wilton_PRL}. In the low temperature regime, the spin and valley resolved longitudinal conductivity is given by
\begin{align}
	\frac{\sigma^{\eta, s}_{xx}(\omega)}{\sigma_0/2 \pi} &= \frac{4 \mu^2 - |\Delta_s^{\eta}|^2}{2 \hbar \mu \Omega} \Theta (2 \mu - |\Delta_s^{\eta}|) \nonumber \\
	&+ \left[ 1 - \frac{|\Delta_s^{\eta}|^2}{\hbar^2 \Omega^2} \right] \tan^{-1} \!\left[ \frac{\hbar \Omega}{M} \right] + \frac{|\Delta_s^{\eta}|^2}{\hbar \Omega M}
\label{sigmaxx}
\end{align}
and the transverse (Hall) conductivity is found to be
\begin{equation}
	\frac{\sigma^{\eta, s}_{xy} (\omega)}{\sigma_0/2 \pi} = \frac{2 \eta \Delta_s^{\eta}}{\hbar \Omega}\tan^{-1} \! \left[ \frac{\hbar \Omega}{M} \right] \,.
\label{sigmaxy}
\end{equation}
In addition, $\sigma^{\eta, s}_{yy} (\omega) = \sigma^{\eta, s}_{xx} (\omega)$ and $\sigma^{\eta, s}_{xy}(\omega) = - \sigma^{\eta, s}_{yx} (\omega)$. The total conductivities are obtained from $\sigma_{ij} (\omega) = \sum_{s, \eta}~\sigma^{\eta, s}_{ij} (\omega)$. In previous equations, $\Theta$ is the Heaviside function, $\sigma_0 = e^2/(4 \pi)$, $M = \text{max}(|\Delta_s^{\eta}|, 2 |\mu|)$, and $\Omega = - i \omega + \Gamma$, with $\Gamma = 1/(2 \tau)$ and $\tau$ being the scattering time that accounts for effects of impurities \cite{Cysne-2016}. It is worth mentioning that we assume the local approximation ($|\vec{q}|\rightarrow 0$) in Eqs. (\ref{sigmaxx}) and (\ref{sigmaxy}). This approximation describes very well the dispersive forces in graphene \cite{localApprox} and should also apply to the other graphene family materials of Table \ref{table2} since they have very similar Fermi velocities.

\begin{figure}[h!]

	\centering
	\includegraphics[width=0.99\linewidth,clip]{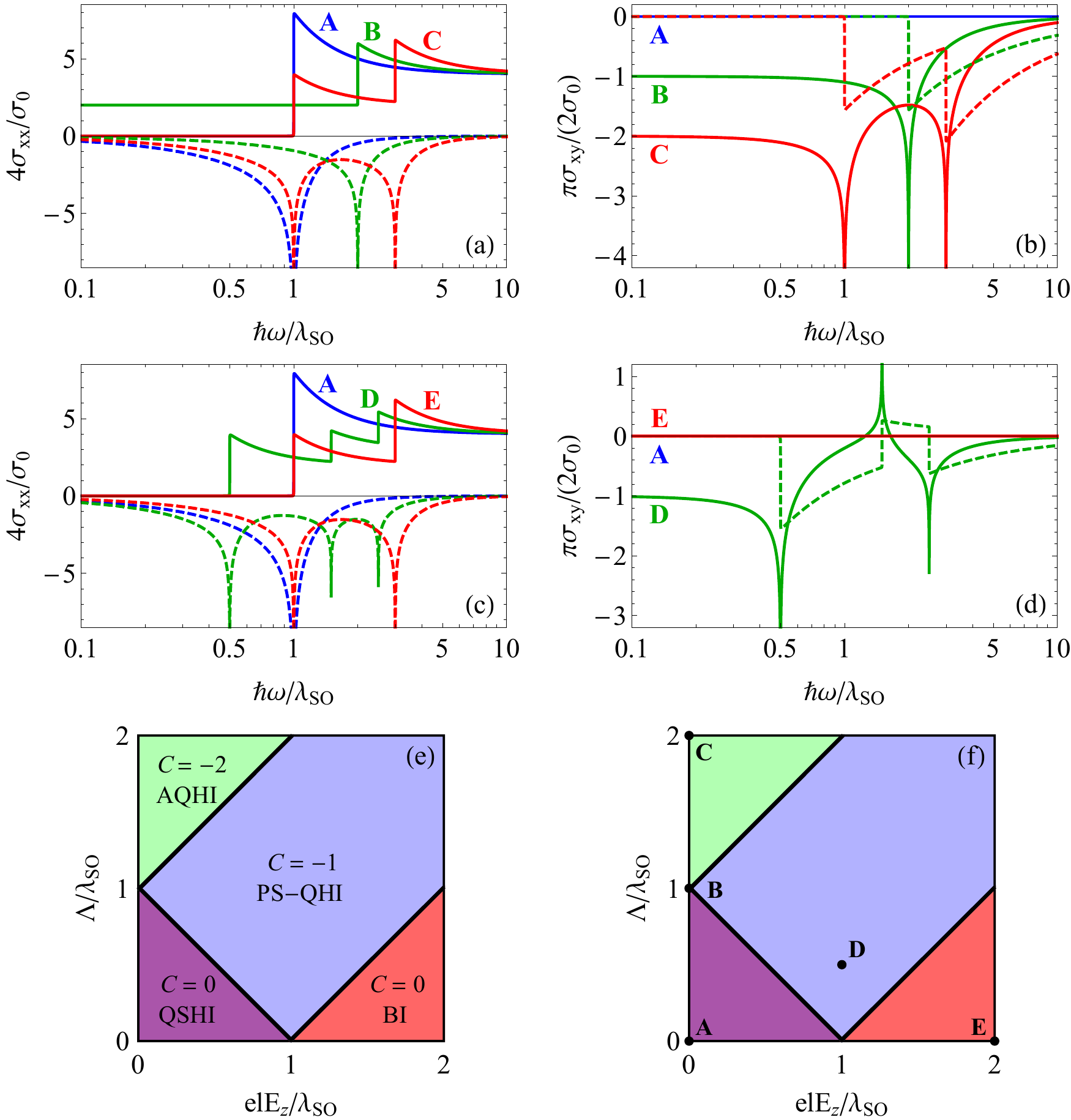}
	\caption{Panels \textbf{(a)} and \textbf{(b)}: Real (solid) and imaginary (dashed) parts of $\sigma_{xx} (\omega)$ and $\sigma_{xy} (\omega)$ for some combinations of parameters $\{ e \ell E_z/\lambda_{\textrm{SO}}, \Lambda/\lambda_{\textrm{SO}}\}$. {\bf A:} $\{0, 0\}$, {\bf B:} $\{0, 1\}$ and {\bf C:} $\{0, 2\}$, as indicated in the topological phase diagram of panel (f). Panels \textbf{(c)} and \textbf{(d)}: the same as before but for {\bf A:} $\{0, 0\}$, {\bf D:} $\{1, 0.5\}$ and {\bf E:} $\{2, 0\}$, as in panel (f). We set $\Gamma = 10^{-4} \lambda_{\textrm{SO}}/\hbar$ and $\mu = 0$ in all cases. Panels \textbf{(e)} and \textbf{(f)}: Topological phase diagram of the graphene family materials described by the Hamiltonian (\ref{Hd}). Panel (e) shows the phase diagram with the acronyms and their respective Chern numbers, while panel (f) sketches the set of parameters used in panels (a)-(d). A detailed discussion of this phase diagram and the optical conductivities can be found in Refs. \cite{Ezawa_2013, grapheneFamily_Nature, grapheneFamily_Wilton_PRL}.}
	\label{FigureA1}

\end{figure}

In Fig. \ref{FigureA1} we plot the real and imaginary parts of the longitudinal and transverse conductivities for different combinations of parameters $\{ \Lambda/\lambda_{\textrm{SO}}, e \ell E_z/\lambda_{\textrm{SO}} \}$, so that we capture some representative points of the topological phase diagram modeled by Eq. (\ref{Hd}). From panels (b) and (d), it can be noticed the relation between the static limit of the transverse conductivity and the Chern number of the topological phase: $\sigma_{xy} (\omega \rightarrow 0) = [e^2/(2h) ] C$ \cite{grapheneFamily_Nature, grapheneFamily_Wilton_PRL, Tse-Niu-2011}. Equations (\ref{sigmaxx}) and (\ref{sigmaxy}) in combination with Eqs. (\ref{rss}-\ref{alpha}) are used in Eq. (\ref{u}) to compute the CP potential between the incident atoms and the 2D graphene family material.

\begin{figure*}[th!]
	\centering
	\includegraphics[width=0.98\linewidth,clip]{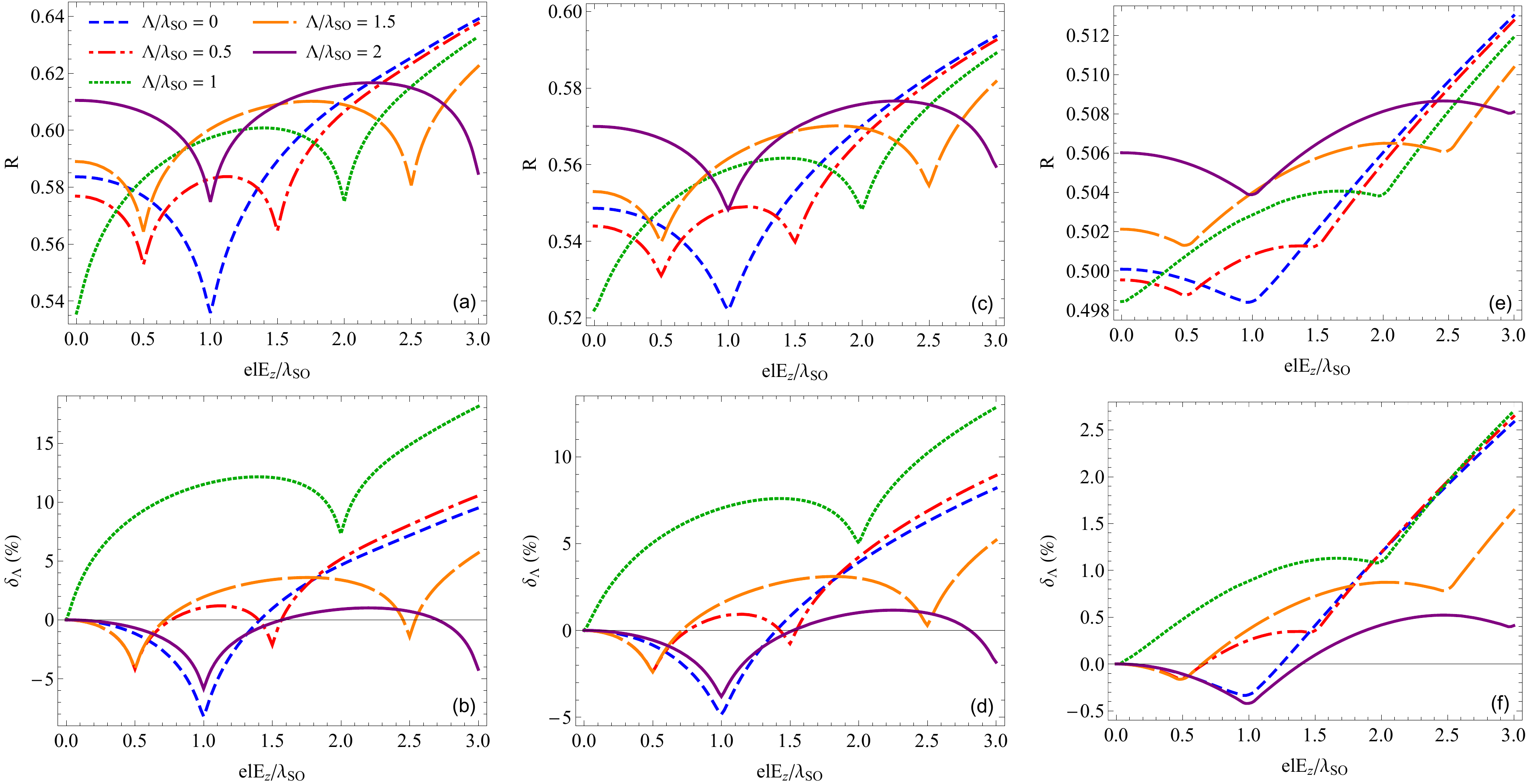}
	\caption{Results for a Na atom with incident energy $E = 10^{-3}$ neV. QR probability as a function of the applied electric field by a suspended sheet of pristine {\bf (a)} stanene, {\bf (c)} germanene and {\bf (e)} silicene. Percentual modification in the QR probability due to the electric field by a suspended sheet of pristine {\bf (b)} stanene, {\bf (d)} germanene and {\bf (f)} silicene. In all cases different curves refer to different values of the laser parameter.}
	\label{FigApp4}
\end{figure*}

\section{Results for Na atom \label{appendC}}

Figure \ref{FigApp4} shows results similar to those presented before, but for QR of a Na atom, which is also relevant in experimental setups \cite{Pasquini-2004, Pasquini-2006}. In this case, we chose an incident energy $E = 10^{-3}$ neV and the outcomes are qualitatively the same as for Rb. The QR probability shows a cusp whenever the graphene family material undergoes a topological phase transition induced by $E_z$ for a fixed $\Lambda$. The degree of tunability of QR probability is also more evident using materials with higher spin-orbit coupling as the reflective material. As we already pointed out, if we compare the degree of control, given by Eq. (\ref{percentual}), for the QR probability of Rb and Na by the same graphene family surface, we find that Rb allows a greater control by $E_z$ than Na. This is due to the fact that Rb is more massive than Na, resulting in a lower energy regime for which the QR of Rb is more tunable than the QR of Na \cite{QR-MagField-2019}. Consequently, by comparing the QR of Rb and Na at the intermediate energy regime, the former is dominated by longer distances (bigger $z_m$) between the atom and the reflective material as compared to the latter, where the CP energy exhibits a higher degree of control with $E_z$.


\end{document}